\documentclass[showpacs,preprintnumbers,amsmath,amssymb,twocolumn,showkeys]{revtex4}
\usepackage{graphicx}   
\usepackage{dcolumn}    
\usepackage{bm}         
\usepackage{ifpdf}
\usepackage{graphicx,amssymb,lineno}
\usepackage{amssymb,lineno,amsfonts}
\usepackage{graphicx}   
\usepackage{dcolumn}    
\usepackage{bm}         
\usepackage{bbm}
\usepackage{mathrsfs}
\usepackage{upgreek}
\usepackage{mathtools}

\begin{document}

\title{Study of Electromagnetically Induced Transparency using 
long-lived Singlet States}

\author{Soumya Singha Roy and T. S. Mahesh}
\email{mahesh.ts@iiserpune.ac.in}
\affiliation{NMR Research Center, \\
Indian Institute of Science Education and Research (IISER), Pune 411008, India}

\date{\today}

\begin{abstract}
{
The long-lived singlet states are useful to study a variety of interesting
quantum phenomena.  In this work we study electromagnetically
induced transparency using a two-qubit system. The singlet state
acts as a `dark state' which does not absorb a probe radiation
in the presence of a control radiation.  Further we
demonstrate that the simultaneous irradiation of probe and control
radiations acts as a dynamical decoupling preserving the singlet
state at higher correlation for longer durations.
}
\end{abstract}

\maketitle

\section*{Introduction}
Simulating quantum phenomena on physical systems usually involves
preparing a pure state, i.e., a definitely known quantum state 
and then carefully controlling its dynamics \cite{chuangbook}.  
Nuclear Magnetic Resonance (NMR) is often regarded 
as a good candidate for studying quantum control \cite{laflammePRA08}.
Initializing an ensemble of nuclear spin systems 
into a pure state is nontrivial.  
Usually an indirect initialization can be achieved and is known as 
a pseudopure state \cite{dieterrev}.  Controlling the nuclear spin dynamics can be
achieved with high fidelity using specially designed radio frequency (RF)
pulses.  
Nuclear spins usually retain 
quantum coherences for durations long enough to carry out
complex unitary evolutions \cite{LevBook}.
Recently it has been established that singlet states
of nuclear spin pairs under certain conditions can
have extraordinarily long lifetimes \cite{LevPRL04,LevittJACS04}.
Singlet life-times often exceed single spin 
coherence times by an order of magnitude or more \cite{sarkar36}
and in a particular case the coherence lasted up to half an hour \cite{LevittJACS08}.  
It has been demonstrated recently that such long-lived nuclear spin states can be utilized
for the preparation of pseudopure states as well as to prepare
high quality Bell states  \cite{maheshpra10}.  

Electromagnetically Induced Transperency (EIT) involves rendering transparent a
small frequency range of an absorption line of a quantum system due
to the distrtuctive interference of transition probabilities induced by
two electromagnetic fields \cite{Harris1990,Harris97}.  
The importance of EIT stems from the fact that it is a purely quantum
mechanical phenomenon and has no classical analogue.
Several applications of EIT arise from the greatly enhanced
non-linear susceptibility in the spectral region of induced
transparency of the medium and the associated steep
dispersion.
In the following we first describe initializing a 2-qubit NMR 
system into a singlet state and 
then describe their applications for studying EIT.

\section*{Initializing Singlet States}
\label{secsingstates}
The Hamiltonian for an ensemble of spin-1/2 nuclear pairs of same isotope, in 
the RF interaction frame, can be expressed as
\begin{eqnarray}
{\cal H^{\mathrm{eff}}} =  h \left[ 
         \frac{\Delta\nu}{2}  I_z^1 
         - \frac{\Delta\nu}{2}I_z^2 
         +  J I^1 \cdot I^2
         +  \nu_{12} I_x^{1,2}
          \right].
\label{heff}         
\end{eqnarray}
Here the RF frequency is assumed to be at the
mean of the two Larmor frequencies, and
$\Delta \nu$, $J$ and $\nu_{12}$ correspond  respectively to the
difference in Larmor frequencies (chemical shift difference), the
scalar coupling constant and the RF amplitude (all in Hz).  
In the limiting case of $\Delta \nu \rightarrow 0$, the system is
said to have magnetic equivalence, and the singlet state 
$\vert S_0 \rangle = (\vert 0 1 \rangle - \vert 1 0 \rangle)/\sqrt{2}$, and
the triplet states $\vert T_1 \rangle = \vert 0 0 \rangle$, 
$\vert T_0 \rangle = (\vert 0 1 \rangle + \vert 1 0 \rangle)/\sqrt{2}$,
and $\vert T_{-1} \rangle = \vert 1 1 \rangle$ 
form an orthonormal eigenbasis of the internal Hamiltonian \cite{LevBook} 
\begin{eqnarray}
{\cal H}_{\mathrm{eq}}^{\mathrm{eff}} = h J I^1 \cdot I^2.
\end{eqnarray}  

\begin{figure}
\includegraphics[width=6cm]{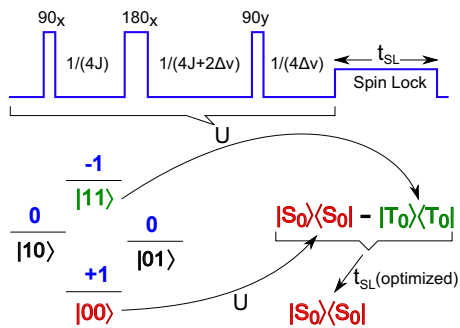} 
\caption{Pulse sequence for preparing singlet state
}
\label{figure1} 
\end{figure}

The state $\vert S_0 \rangle \langle S_0 \vert -\vert T_0 \rangle \langle T_0 \vert$
can be easily prepared from the equilibrium density matrix $I_z^1 + I_z^2$ 
by using the pulse sequence shown in Figure \ref{figure1}.
The initial state is of the form
$\vert 00 \rangle \langle 00 \vert - \vert 11 \rangle \langle 11 \vert$.
Let U be the propagator for the pulse sequence.
It can be easily seen that $U \vert 00 \rangle = \vert S_0 \rangle$
and $U \vert 11 \rangle = \vert T_0 \rangle$ up to overall phases.  
Thus the pulse sequence in Figure \ref{figure1} prepares a non-coherent
statistical mixture of states $\vert S_0 \rangle$ and $\vert T_0 \rangle$.

The equivalence Hamiltonian may be realized
by using a spin-lock \cite{LevittJACS04,BodenCPC07}.
Levitt and co-workers demonstrated that under 
${\cal H}_{\mathrm{eq}}^{\mathrm{eff}}$
the decay constant of singlet state $\vert S_0 \rangle$ is much larger than $T_1$, 
and hence called it a long-lived state \cite{LevPRL04,LevittJACS04}.  
Detailed theoretical analysis of singlet state decay have  been
provided by Levitt and co-workers \cite{LevittJCP05,LevJCP09} and by 
Karthik and Bodenhausen \cite{BodenJMR06}.
Recently we had reported that high-fidelity singlet states
can be easily prepared and characterized in two-spin 
systems under various experimental conditions \cite{maheshjmr10}.  

Due to the long lifetime of the singlet state, only 
$\vert S_0 \rangle$ state survives after a sufficiently long spin-lock
as illustrated in Figure 1
\cite{maheshjmr10}.
Based on this phenomenon, we had proposed that the singlet state can
be used to efficiently initialize NMR quantum registers \cite{maheshpra10}.
The pseudopure singlet state thus prepared can be used for further experiments on 
quantum information.  In the following we describe the experimental
demonstration of EIT using singlet states the dark state.

\section*{Electromagnetically Induced Transparency}
\begin{figure}
\includegraphics[width=7cm]{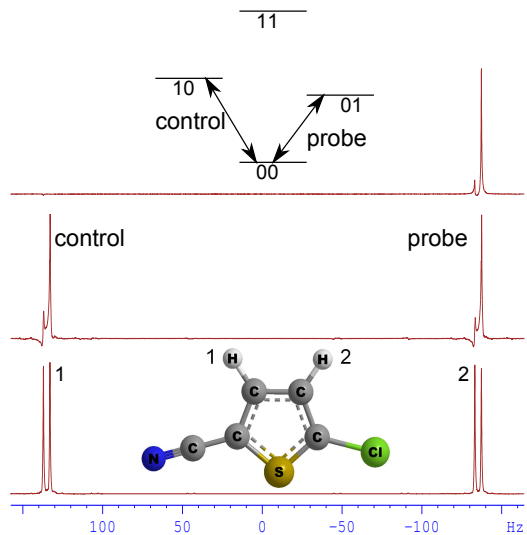} 
\caption{
The energy level diagram indicates the three states of the 2-qubit system
coupled by two electromagnetic radiations.
The upper trace is spectrum obtained by applying only 
the probe field on the equilibrium state.  The middle trace
is that obtained by applying the probe and the control fields
simultaneously.  The bottom trace is the reference spectrum for
the compound.  Molecular structure of 5-chlorothiophene-2-carbonitrile
is shown in the inset.
}
\label{figure2} 
\end{figure}

EIT involves three states of a quantum system coupled
by two electromagnetic radiations.  If the system 
 $(\vert 01 \rangle - \vert 10 \rangle)/\sqrt{2}$, 
 then it is a stationary
state under the two radiations and no net absorption or emission 
takes place \cite{Harris1990,Harris97}.  
One of the two fields is known as `control'
and the other as `probe' (see Figure \ref{figure2}).
The stationary coherent state of the system under these two
radiations is known as the `dark state'.

EIT has been studied extensively in optical systems and has
several applications such as enhancement of non-linear optical frequency
conversion \cite{Harris1990}, resonant four wave mixing \cite{Hemmer},
and coherent preparation of a maximal coherence \cite{Jain}.
More recently, exciting applications in quantum information processing 
have also been demonstrated \cite{spiller,filippo}.
The group velocity of photons carrying quantum information can be 
made vanishing inside an EIT medium \cite{Fleischhauer}.  

EIT has also been studied using NMR.  Phase 
coherence of EIT has been studied by Murali et al using spin 7/2 nuclei of
$^{133}$Cs system partially oriented in a liquid crystalline medium \cite{muraliEit}.
Cory and co-workers used a liquid state NMR system to demonstrated EIT
\cite{coryEIT}. These works involved an initial preparation of a 
longitudinal pseudopure
state which was then converted into a dark state using a sequence of 
RF pulses.  

In our demonstration using the pulse sequence in Figure \ref{figure1}, 
the dark state is naturally prepared because
of the long-lived nature of the singlet state.  After obtaining the
maximum fidelity of the dark state, the equivalence Hamiltonian (spin-lock) is
turned off and the probe and the control fields are applied simultaneously as 
shown in Figure \ref{figure2}. 
Further, in our implementation we have carried out the density matrix tomography 
of the dark states which enables
us to quantitatively monitor the dynamics of the dark state.

Three levels 
$\{\vert 00 \rangle, \vert 01 \rangle, \vert 10 \rangle \}$
of the two $^1$H spins of 
5-chlorothiophene-2-carbonitrile 
were used to study EIT phenomenon (Figure \ref{figure2}).  
About 5 mg of the sample was dissolved in 0.75 ml of dimethyl sulphoxide-D6 and
all the experiments are carried out in a 
Bruker 500 MHz NMR spectrometer at 300 K.  The chemical shifts difference
between these spins are 270.3 Hz and the scalar coupling is 4.1 Hz.
The spin lattice relaxation time constant for the two spins obtained from 
inversion recovery experiment are about 6.3 s.
RF spin-lock was achieved by WALTZ-16 modulations at 2 kHz amplitude.
The lifetime of the singlet state was about 12.0 s, about 
two times the $T_1$ values of the individual spins implying the 
long-lived nature of the singlet state. 
The quality of the singlet state was estimated by the density matrix
tomography \cite{maheshjmr10}.  
The spin-lock duration of 15 s provided a
correlation of about 0.991 with the singlet state.

As shown in Figure \ref{figure2}, the field on 
$\vert 00 \rangle \leftrightarrow \vert 01 \rangle$ transition
is labeled as probe and that on
$\vert 00 \rangle \leftrightarrow \vert 10 \rangle$ transition 
is labeled as the control.
To study the effect of only the probe field, a simple Gaussian
pulse of duration 520 ms was sufficient.  The excitation 
spectrum of the Gaussian pulse is shown in the upper
trace of Figure \ref{figure2}.
The simultaneous irradiation of the probe and the control 
fields were achieved by a strongly modulated pulse of 
duration 240 ms, which was designed to be robust against
the spatial inhomogeneity of RF amplitude
\cite{fortunato,maheshqnge}. Repeated application of this
pulse allowed us to monitor the dark state over extended
durations of time.
The excitation spectrum of the simultaneous irradiation of the probe and the 
control is shown in the middle trace of Figure \ref{figure2}.
The reference spectrum is in the lower trace of Figure \ref{figure2}.
Of course, the selectivity of the probe and the control 
are not perfect.  There are small amplitudes for the transitions 
connecting
the $\vert 11 \rangle$ level, which should ideally have been left 
isolated.


\begin{figure}
\includegraphics[width=8.7cm]{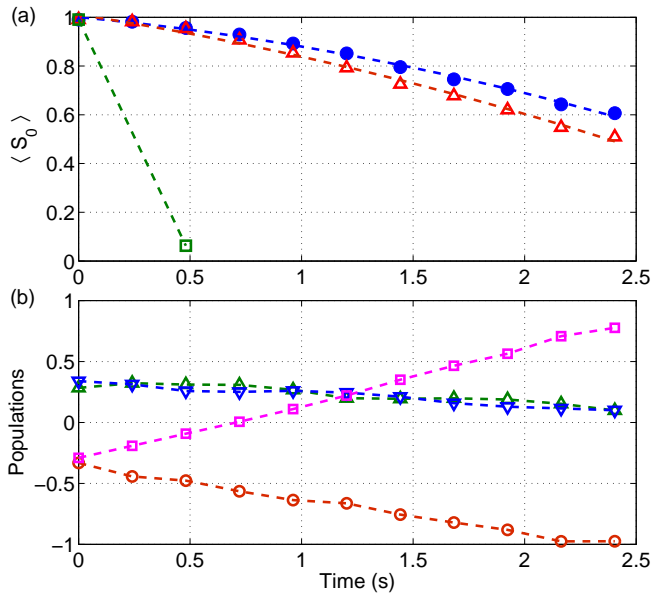} 
\caption{(a) The correlation of the experimental dark state state
under (i) no irradiation (triangles), (ii) under only the probe
field (squares), and (iii) under the combined irradiation of
probe and control pulses (filled circles).
(b) The deviation populations of 
$\vert 00 \rangle$ (circles),
$\vert 01 \rangle$ (down-triangle),
$\vert 10 \rangle$ (up-triangle), and
$\vert 11 \rangle$ (squares).
The sum of all the deviation populations is zero as expected
for a deviation density matrix.}
\label{figure3} 
\end{figure}

We have monitored the state of the system
at regular intervals using density matrix tomography.  Figure \ref{figure3}a
shows the correlation 
\begin{eqnarray}
\langle S_0 \rangle = \frac{\mathrm{trace}(\rho_\mathrm{exp} \cdot \rho_\mathrm{th})}
       {\sqrt{{\mathrm{trace}(\rho_\mathrm{exp}^2)}\;{\mathrm{trace}(\rho_\mathrm{th}^2)}}}.
\end{eqnarray}
of the experimental 
dark state $\rho_\mathrm{exp}$ with the theoretical singlet
state $\rho_\mathrm{th} = \vert S_0 \rangle \langle S_0 \vert$.

Under the application of only control field, the dark state
immediately collapses and the correlation of the experimental density 
matrix drops immediately.  The dynamics in this case
is expected to be oscillatory, and we have not studied this situation
with further tomography.
Under the simultaneous application of the probe and control fields, the
dark state is maintained for much longer time.  Ideally
the dark state is a stationary state under the combined irradiation.
But the decay of the experimental dark state is due to the non-ideality
of the control fields
(limited selectivity and RF inhomogeneity)
and due to decoherence of the dark state.

Figure 3 also shows the decay of the dark state under no irradiation.
In this case the decay is slightly faster than that of the combined
irradiation.  Thus the combined irradiation of the probe and control
fields acts as a dynamical decoupling of the dark state from the
environment, storing it for longer duration of time \cite{viola99,maheshudd}.

The density matrix tomography allows us to monitor the populations
at various levels during the simultaneous application of probe and control
\cite{maheshjmr10}.  
As shown in Figure \ref{figure3}b, the populations of 
$\vert 01 \rangle$ and $\vert 10 \rangle$ remain equal and almost
constant for up to two seconds, thus demonstrating the phenomenon 
of coherent population trapping.  However, there appears to be a slow exchange
populations between $\vert 00 \rangle$ and $\vert 11 \rangle$.  This
exchange is induced by the imperfect selectivity of the
probe and control pulses.

\begin{figure}
\includegraphics[width=6.7cm,angle=-90]{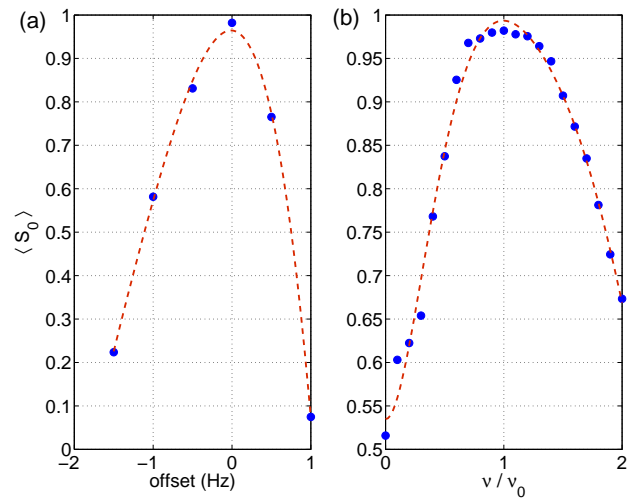} 
\caption{The correlation of the experimental dark state state
under under the simultaneous irradiation of probe and control fields
as a function of (a) offset and (b) relative amplitude of the probe
irradiation.  In (b) $\nu$ and $\nu_0$ are the probe and control
amplitudes.  The duration of RF was 240 ms in all the experiments.
}
\label{figure4} 
\end{figure}

We have also studied the dependence of the dark state correlation 
on the resonance offset of probe and on
relative amplitudes of probe and control.  These results are displayed
in Figures \ref{figure4}a and \ref{figure4}b.  It can be seen that
the correlation is very sensitive to the offset of probe (and control).
Correlation drops to below 0.1 by increasing the offset even by
1 Hz as shown in Figure \ref{figure4}a.  Similarly from Figure \ref{figure4}b,
it can be seen that the correlation is maximum when the amplitudes of the
probe and control fields are equal.

\section*{Conclusions}
Due to the long-lived nature, nuclear spin-pair
ensembles can be conveniently initialized to singlet states
with high fidelity.
Treating the singlet state as a dark state
under the simultaneous application of probe and control fields,
we demonstrated electromagnetically induced transparency, which
is a purely quantum mechanical phenomenon.  It is shown experimentally 
that the dark state is retained for fairly long durations up to
about two seconds.  Monitoring the populations of various levels
demonstrated the coherent population trapping and also 
indicated the effect of decoherence and of 
non-selectivity of the control and probe fields.
Finally we have studied the sensitivity of the dark state
to the offset and the relative amplitude of the probe with
respect to the control field.

\acknowledgments
Authors gratefully acknowledge discussions with Prof. Anil Kumar, Dr. Karthik
Gopalakrishnan, and Prof. G. S. Agarwal.  
The use of 500 MHz NMR spectrometer at NMR Research Center, IISER-Pune
is also acknowledged.

\end{document}